\documentclass[
 twocolumn,noshowpacs,noshowkeys,pra,aps,
 amsmath,amssymb,10pt,
]{revtex4-1}
\usepackage{graphicx}
\usepackage{dcolumn}
\usepackage{bm}
\usepackage[utf8]{inputenc}
\usepackage[T1]{fontenc}
\usepackage{mathptmx}
\usepackage{etoolbox}

\begin{document}

\title[]{Characterization of Long-Term Stable Photonic Microwaves\\
based on a Difference Frequency Comb}

\author{S. Mueller}
 \altaffiliation[Also at ]{Institute of Microwaves and Photonics, Friedrich-Alexander University, Erlangen, Germany }
\author{T. Puppe}%
\affiliation{ 
TOPTICA Photonics AG, Graefelfing, Germany
}
\date{April 16, 2025}
\textbf{}
\begin{abstract}
We report on a novel method for optical microwave generation using a frequency comb based on difference-frequency generation, which passively eliminates the carrier-envelope offset frequency ($f_{\mathrm{ceo}}$), with the repetition rate ($f_{\mathrm{rep}}$) locked to an optical reference. We demonstrate the generation of ultra-low phase noise microwave signals by transferring the stability of the optical reference to 9.6\,GHz, reaching noise levels of -147\,dBc\,/\,Hz at 1\,kHz offset. The optimization of pulse timing after interleaving and a scheme for additional long-term stabilization of the microwave signal to GPS standards are discussed. This work presents a new variant of highly stable RF signal generation for precision applications, such as radar, atomic clock local oscillators and optical quantum technologies.
\end{abstract}

\maketitle

\section{Introduction}
Low-noise and low timing jitter radio frequency (RF) signals in the range of a few GHz to several hundred GHz are of increasing relevance in a variety of technical fields. Among other applications, radar \cite{Ghelfi2014}, synchronization of very long interferometer paths \cite{Tomio2024}, atomic clock local oscillators \cite{Lipphardt2023}, and the generation and analysis of high-frequency signals - in particular within the context of 5G and 6G communications \cite{Santacruz2021, Delmade2023} - greatly benefit from microwave oscillators with very low phase noise. Additionally, the long-term stability of the generated signal and potential referencing to an absolute frequency standard are critical for some of these use cases \cite{Lipphardt2023}. In the last decade, RF signals based on optical frequency division have shown the potential of generating lower phase noise than conventional electronic oscillators, enabling and advancing ultra-low noise and high-frequency applications \cite{Zobel2019}. \\
\indent Optical frequency division microwaves (OFDµW) are typically based on a frequency comb with stable carrier-envelope offset frequency ($f_{\mathrm{ceo}}$), which is locked to a reference laser. The most widely reported method in the literature detects and locks the $f_{\mathrm{ceo}}$ using an f-2f interferometer, while a beat between one of the optical comb lines and a continuous wave (CW) ultra-low noise (ULN) laser is used to stabilize the comb repetition rate ($f_{\mathrm{rep}}$) \cite{Fortier2011}. Within the locking bandwidth, the phase noise of any optical comb tooth is given by the scaled ULN laser noise \cite{Puppe2016} and residual $f_{\mathrm{ceo}}$ noise. At higher offset frequencies, the noise corresponds to that of the free-running comb. By converting a harmonic of the repetition rate of a mode-locked laser using a fast, highly linear photodiode, electronic RF signals with phase noise given by the scaled phase noise of the repetition rate $f_{\mathrm{rep}}$, are generated. \cite{Xie2017} \\
\indent As an alternative to detecting and locking $f_{\mathrm{ceo}}$, a comb based on difference frequency generation (DFG) passively eliminates the $f_{\mathrm{ceo}}$ from the spectrum, leaving the repetition rate as the only free parameter. This work illustrates the results of optical microwave generation based on a DFG comb, which removes the potential for crosstalk between the usual two stabilization loops for $f_{\mathrm{ceo}}$ and $f_{\mathrm{rep}}$. Specifically, we discuss the noise characteristics and spectral purity of a microwave at 9.6\,GHz, improved SNR by implementation of an optical pulse interleaver, and a way to reference the microwave signal to a GPS-disciplined RF frequency standard. 

\section{Setup}

\begin{figure}[b]
\includegraphics[width=\linewidth]{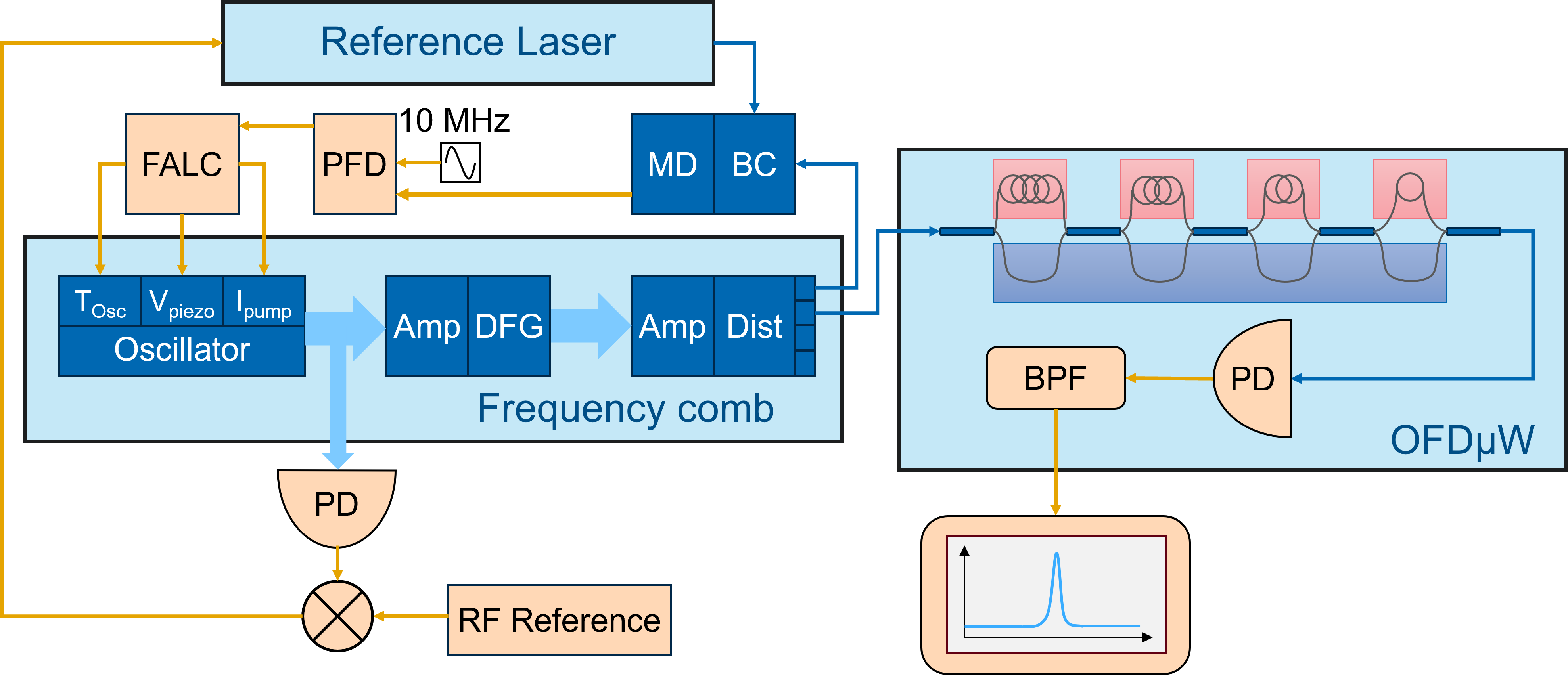}
\label{fig_Setup_SingleSystem}
\caption{ The optical microwave setup consists of a reference laser, the comb based on difference frequency generation (DFC CORE 200 +), a pulse interleaver and a modified uni-travelling carrier photodiode (PD) with 9.6\,GHz RF bandpass filter (BPF). MD: Monochromatic detector (bandpass + photodiode) BC: Beam combiner, FALC: Fast analog laser controller, PFD: Phase frequency detector}
\end{figure}

\begin{figure*}
\includegraphics[width=0.7\linewidth]{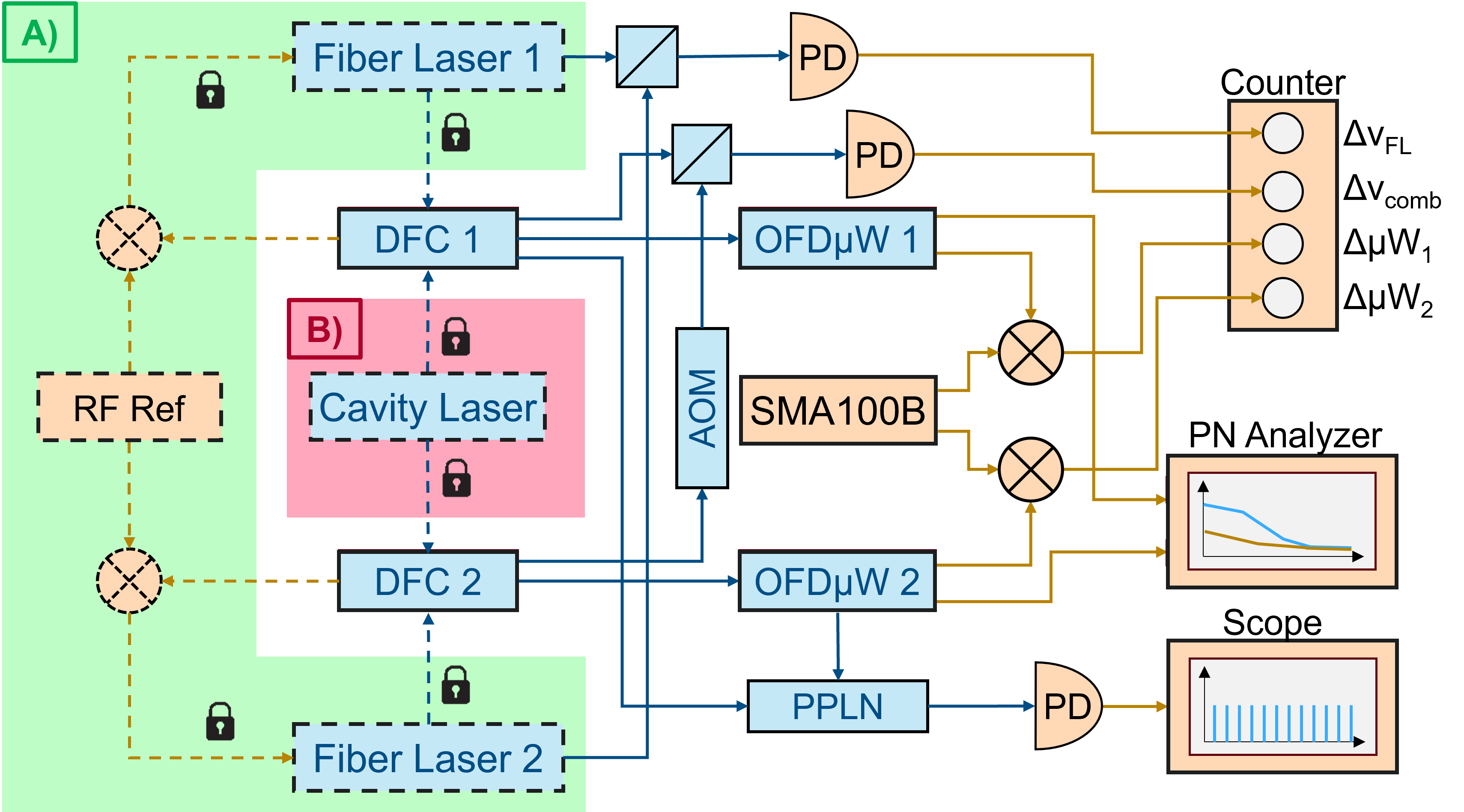}
\caption{To characterize the OFDµW signals with increased sensitivity, two identical systems are measured against each other. This is done for the microwave signals, as well as the optical beat between both combs. For long-term stability analysis, a frequency counter (K+K FXE) simultaneously records the microwave signals mixed to 100\,MHz with a common local oscillator (SMA100B) as well as the optical comb beat and beat between the two fiber lasers. A periodically-poled $\mathrm{LiNbO_3}$ crystal (PPLN) is used to analyze the pulse timing via cross-correlation of the pulse train before and after the pulse interleaver. DFC: Difference frequency comb, AOM: acusto-optical modulator, OFDµW: Optical frequency division microwave consisting of interleaver and optical-to-RF conversion. A) The two combs are optically locked to two independent fiber lasers, which in turn are referenced to a common GPS disciplined RF signal. B) Both combs are locked to a common cavity-locked diode laser. \label{fig_Setup_TwoSystems}}
\end{figure*}

Our setup to generate photonic microwaves is based on a DFG-comb (TOPTICA DFC CORE 200 +). The frequency comb consists of a mode-locked erbium-fiber oscillator with a repetition rate of $f_{\mathrm{rep}}$ = 200\,MHz, a nonlinear amplifier broadening the optical spectrum to 850 and 1860\,nm, the DFG unit, generating $f_{\mathrm{ceo}}$-free comb modes around 1560\,nm \cite{Kliese2016}, and an output amplifier providing up to 8 identical output ports. Each of the ports yields approximately 13\,mW average power. One output is amplified to 60\,mW to drive the interleaver and opto-electronic conversion. The frequencies of DFG-comb modes are given by the simplified comb equation \cite{Benkler2005}:

\begin{equation}
 f_n = f_{\mathrm{rep}} \times n + (f_{ceo} \equiv 0) = f_{\mathrm{rep}} \times n.
\label{equation1_SimplifiedComb}
\end{equation}

\indent For the optoelectronic conversion of the OFDµW, a commercial, highly linear modified uni-traveling carrier (MUTC) photodiode is used (Coherent VPDV2120). To increase the carrier power at specific harmonics of $f_{\mathrm{rep}}$ while preventing saturation of the photodiode, the effective comb repetition rate is increased from 200\,MHz to 3.2\,GHz utilizing an all-fiber 4-stage pulse interleaver. Each interleaver stage is built as an asymmetric Mach-Zehnder interferometer, with an optical path length difference corresponding to half the pulse spacing at the stage input. Precision splicing of the stages allows to reach the optimal path length difference - and therefore pulse spacing - by fine tuning the temperature. With an optimized interleaver, RF levels of up to 0 dBm (-3 dBm behind an RF bandpass filter) at 9.6\,GHz are generated. The microwave signal is characterized using a Rohde \& Schwarz FSWP26 phase noise analyzer. A schematic of the comb and microwave setup is shown in Fig~1. \\
\indent The OFDµW is stabilized with a phase-locked loop (PLL) to an optical reference (see Fig. 1). A beam combiner, optical bandpass filter, and photodiode are used to generate a high SNR beat of the CW reference laser with a comb mode. The RF beat is phase-locked to an RF local oscillator derived from a 10\,MHz reference using a phase frequency detector and a fast analog loop filter. The locking bandwidth is >\,450\,kHz, where the fast fluctuations are addressed by the oscillator pump current, while a piezo as well as the oscillator temperature are used to compensate for acoustic disturbances and long-term drifts, respectively. \\
\indent As optical reference, two types of lasers are used alternatively, a cost-efficient and compact low-noise fiber laser (NKT Koheras BASIK X15) (Fig. 2A), and an ULN laser consisting of a diode laser locked to a high-finesse Fabry-Pérot cavity based on ultra-low expansion glass (modified TOPTICA CLS) (Fig. 2B). The fiber laser can be modulated via an analog input, to change the wavelength and establish long-term stabilization \cite{Kliese2021}. When $f_{\mathrm{rep}}$ of the comb is locked to the cavity-locked ULN laser, the OFDµW phase noise at low Fourier frequencies can not be resolved in reasonable measurement times using cross-correlation measurements with respect to the internal RF local oscillators. Hence, a second, identical system is used as an external local oscillator for the FSWP to measure the relative phase noise compared to the first system as device under test (DUT). As a more cost-effective and compact alternative to the cavity-locked ULN laser, the frequency comb can be locked to the low-noise fiber laser. The setup allows for the generation of the phase error between the optically locked comb and a GPS-stabilized oven-controlled crystal RF reference (OCXO). The $\mathrm{10^{th}}$ harmonic of the 80\,MHz fundamental OCXO frequency at 800\,MHz is used to generate the phase error with respect to the $\mathrm{4^{th}}$ harmonic of the comb repetition rate. A control signal based on this error is used to stabilize the wavelength of the CW fiber laser, thus referencing fiber laser and the microwave signal to GPS and allowing for absolute stability on longer time scales (see~Fig.~2A). \\
\indent To characterize the OFDµW long-term stability, the signals at 9.6\,GHz are mixed to an intermediate frequency (IF) of 100\,MHz with a local oscillator provided by a signal generator (Rohde \& Schwarz SMA100B) at 9.5\,GHz, and recorded with a frequency counter (K+K FXE). A beat between the two frequency combs, generated by frequency shifting DFC 2 by 110.7\,MHz with an acousto-optical modulator (AOM) and establishing a beat with DFC 1, is counted synchronously at 200.0\,MHz - 110.7\,MHz = 89.3\,MHz. \\
\indent Lastly, the pulse timing after the OFDµW 2 interleaver is measured by cross-correlating the pulse train with the fundamental pulse train of DFC 1 using a periodically-poled $\mathrm{LiNbO_3}$ crystal (PPLN) with orthogonal polarizations (see next section \ref{PulseTiming}).

\section{Results}

\subsection{Pulse timing}
\label{PulseTiming}

For short-pulsed illumination of photodiodes, the signal shot noise is shifted mainly into the amplitude quadrature, while the direct contribution to phase noise can be suppressed by several orders of magnitude \cite{Quinlan2013}. This effect is reduced, however, when deviations in the interleaver path lengths lead to imperfect pulse spacing \cite{Quinlan2014}. Therefore, the timing of the pulses behind the four-stage interleaver is characterized. The interleaver delays are optimized by fine tuning the temperature of one interferometer arm for each respective stage. \\
\begin{figure}[t]
\includegraphics[width=\linewidth]{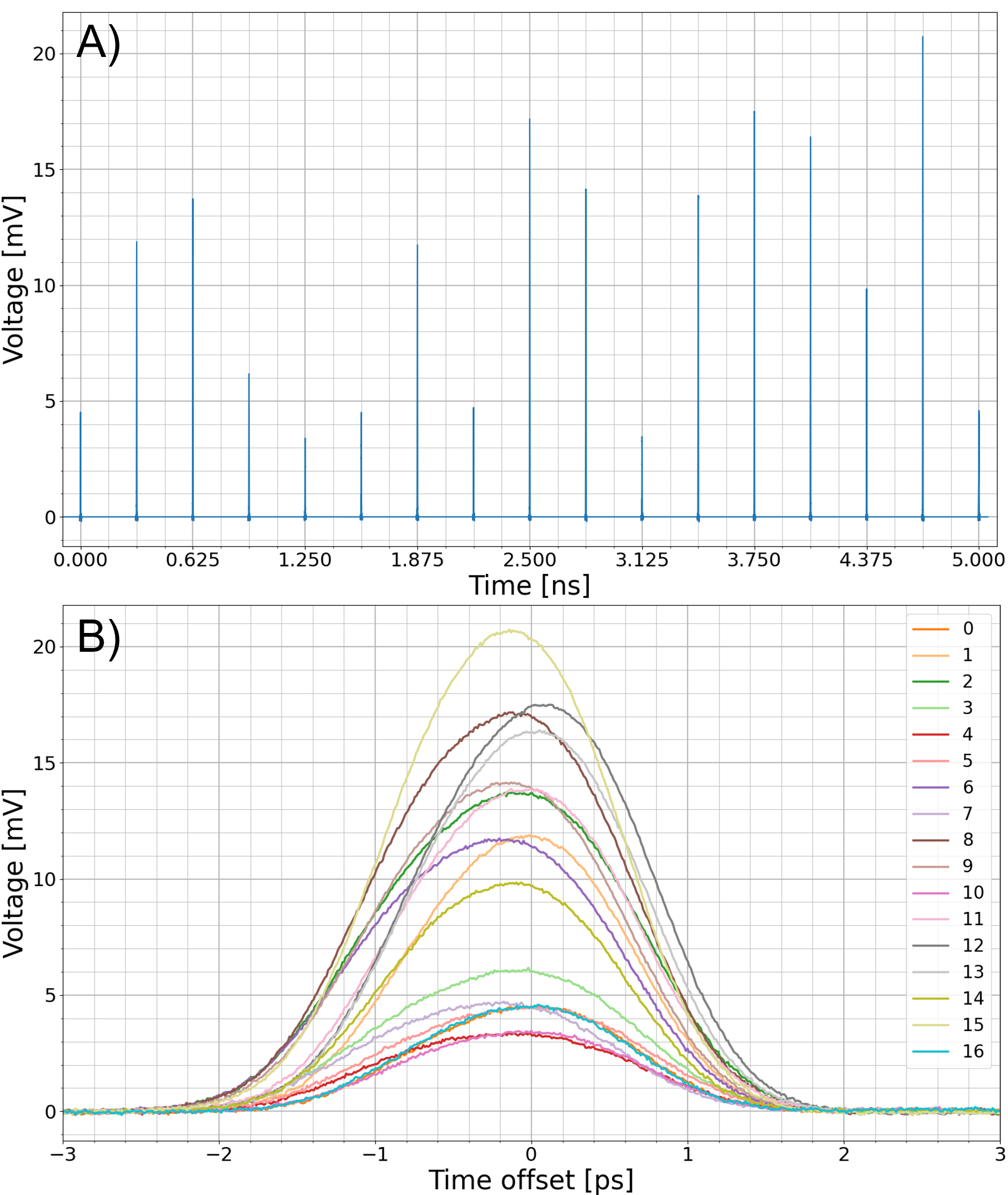}
\caption{A): Cross-correlation time trace of the 16 pulses generated from the 4-stage interleaver, including the first pulse of the next cycle. Amplitude variation is due to the polarization sensitivity of the nonlinear crystal and does not represent pulse energy B) Pulse time offset from their respective ideal timing position n\,/\,(16$ \times f_{\mathrm{rep}}$). \label{fig_PulsesAndTimeTrace_vertical}}
\end{figure}
\indent For the timing analysis, the interleaved pulse train of one microwave system is overlapped with the comb output of the second system to generate a background-free pulse cross-correlation using a PPLN with orthogonal polarizations. The local oscillator (LO) used for the OPLL of the second comb is tuned by 10\,MHz, thereby slightly shifting the comb repetition rate by approximately 10\,Hz. Hence, the pulses which are asynchronously sampled at the crystal output can be detected with a photodiode and analyzed with a scope. The average of 100 traces of the 16 pulses, including the repetition of the first pulse, is shown in Fig.\,\ref{fig_PulsesAndTimeTrace_vertical} A). The inverse pulse spacing is given by the repetition rate difference $\Delta f_{\mathrm{rep}}$ of the two combs, which is estimated by scaling the offset in the optical domain to the effective repetition rate:

\begin{equation}
\Delta f_{\mathrm{rep}} = \frac{3.200 \ \mathrm{GHz}}{194.410 \ \mathrm{THz}} \times 10 \ \mathrm{MHz} = 164.6 \ \mathrm{Hz}.
\end{equation}
\vspace{1mm} 

The difference in pulse amplitude is resulting from the fact that single-mode non polarization maintaining fiber components are used for the interleaver, while the cross-correlation is sensitive to polarization in the frequency doubling process. Because the crystal of a balanced optical cross-correlator setup shows significant difference in dispersion on the polarization axes, the measured signals are expected to be considerably longer than the real pulse duration. To verify this, the optical pulses were characterized with an auto-correlator after propagating through a fiber with the length of the average path in the interleaver yielding a full width at half maximum duration of approximately 230\,fs. Fig.\,\ref{fig_PulsesAndTimeTrace_vertical} B) shows the offset of the pulses compared to their ideal timing according to the estimation of $\Delta f_{\mathrm{rep}}$. Here, pulse number 16 corresponds to the first pulse of the next fundamental pulse after 5 ns and propagates through the same optical path as pulse number 0. The RMS value for the pulse center of gravity deviation of 155 fs is estimated by fitting a gaussian function to each of the pulses and retrieving the offset. It should be noted that this offset for each pulse remains at a constant value and does not directly influence the microwave phase noise. Instead, it only contributes to the residual RF power in the suppressed harmonics of $f_{\mathrm{rep}}$.

\subsection{Phase noise transfer}
\label{phase noise transfer}

\begin{figure}[t]
\includegraphics[width=\linewidth]{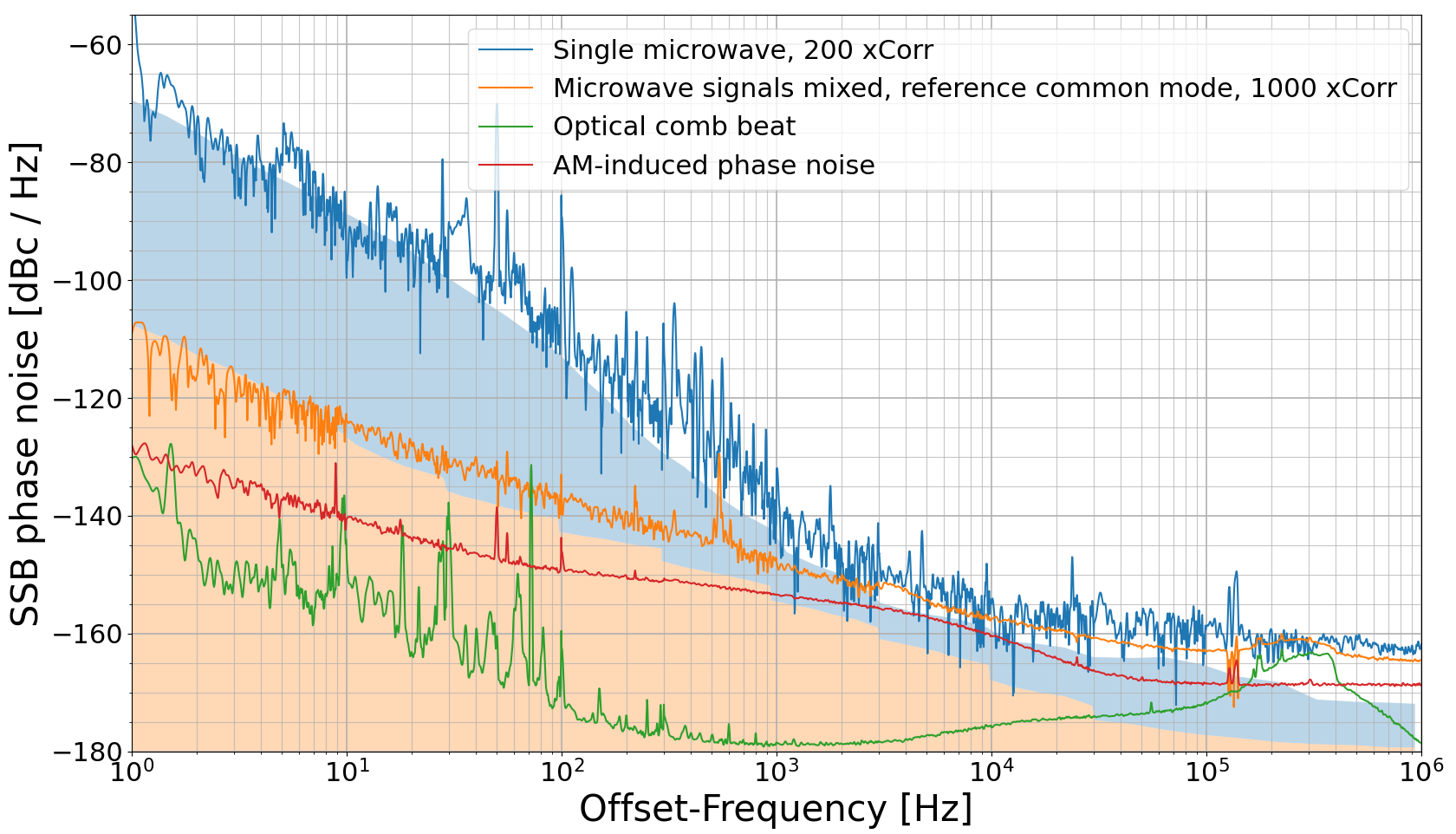}
\caption{Phase noise measurement of a single optical microwave signal at 9.6\,GHz (blue), locked to the CLS, limited by the cross-correlation measurement time at low offset frequencies, as indicated by the shaded area. The comparison of both microwaves against each other (orange) as well as the scaled phase noise on the optical beat between the frequency combs (green).}
\label{fig_Results_PhaseNoise}
\end{figure}

In case $f_{\mathrm{rep}}$ is locked to the cavity-locked ULN laser, the OFD microwave phase noise measurement is limited by the resolution of the FSWP for reasonable cross-correlation measurement times. The single-sideband phase noise of a single system (200 cross-correlations) is shown in Fig.\,\ref{fig_Results_PhaseNoise} together with the residual measurement of two systems, both locked to the same optical reference. The cross-correlation gain indicator of the FSWP \cite{Feldhaus2016} is represented by the shaded region for each trace and the respective number of cross-correlations. The number of correlations is specified for the measurement at 1\,Hz; for higher Fourier frequencies, the number of correlations is higher, as the measurement time needed is lower. \\
\indent For the signal power of 0 dBm generated directly from the photodiode, the Johnson-Nyquist noise is expected to be \mbox{-174\,dBc\,/\,Hz} with the direct contribution to the phase noise being half of that \mbox{(-177\,dBc\,/\,Hz)}. For a train of short pulses, most of the shot noise is contained in the amplitude and only a small part is shifted to the phase quadrature \cite{Quinlan2013}. Based on the results of section \ref{PulseTiming}, we estimate the direct shot noise contribution to the signal PN as \mbox{-212\,dBc\,/\,Hz}, using the calculation established by Quinlan et al. \cite{Quinlan2013}:

\begin{eqnarray}\label{lphi}
&L_{\Phi, sh} = 10 \, log_{10}(\frac{q I_{avg} R |H_n(f_{rep})|^2}{P(f_{rep})} \times (1 - e^{-(2 \pi f_r \tau)^2})) \\
&= 10  \, log_{10}(\frac{1.602 \times 10^{-19} \mathrm{C} \times 13.6 \times 10^{-3} \mathrm{A} \times 50 \ \Omega \times 0.5^2 }{ 10^{-3} \mathrm{W}} \nonumber \\
&\times (1-exp(-(2 \pi \times 9.6 \ \mathrm{GHz} \times 230 \ \mathrm{fs})^2))) \nonumber
\end{eqnarray}
\vspace{1mm} 

Lastly, the amplitude-to-phase-conversion coefficient for this mode of operation is measured to be 23\,dB. The AM-induced phase noise is shown in Fig.\,\ref{fig_Results_PhaseNoise} and poses a limit on the phase noise to -168\,dBc\,/\,Hz. While at some frequencies, the noise contributions of the phase locked loops might become important, the remaining limitations at other frequencies warrant further investigation. For the common mode measurement, the noise approaches a 1\,/\,f noise at Fourier offsets below 10\,kHz at levels comparable to the results which other groups have identified as flicker noise in MUTC diodes \cite{Fortier2013, Lee2021}. For offsets below 10\,Hz, the measurement is potentially limited by the remaining noise floor after 1000 cross-correlations over approximately 10 hours of measurement time. 

\subsection{Long-term stability}
Some applications for low-noise RF oscillators require not only excellent phase noise at short time scales, but also good long-term stability or absolute frequency reference to standard time. The frequency combs are locked to separate optical references as before. By using the phase error between comb repetition rate and an RF reference oscillator at the fourth harmonic of $f_{\mathrm{rep}}$ at 800\,MHz, an additional control loop absolutely stabilizes the wavelength of the CW lasers and thus $f_{\mathrm{rep}}$ (see Eq.\,\ref{equation1_SimplifiedComb}). The RF oscillator is locked to a 10\,MHz reference signal which in turn is derived from a GPS-disciplined OCXO. As shown in Fig.\,\ref{fig_Setup_TwoSystems}, the microwave signals are mixed with a common local oscillator at 9.5\,GHz to an IF of 100\,MHz to be counted.\\
\begin{figure}[t]
\includegraphics[width=\linewidth]{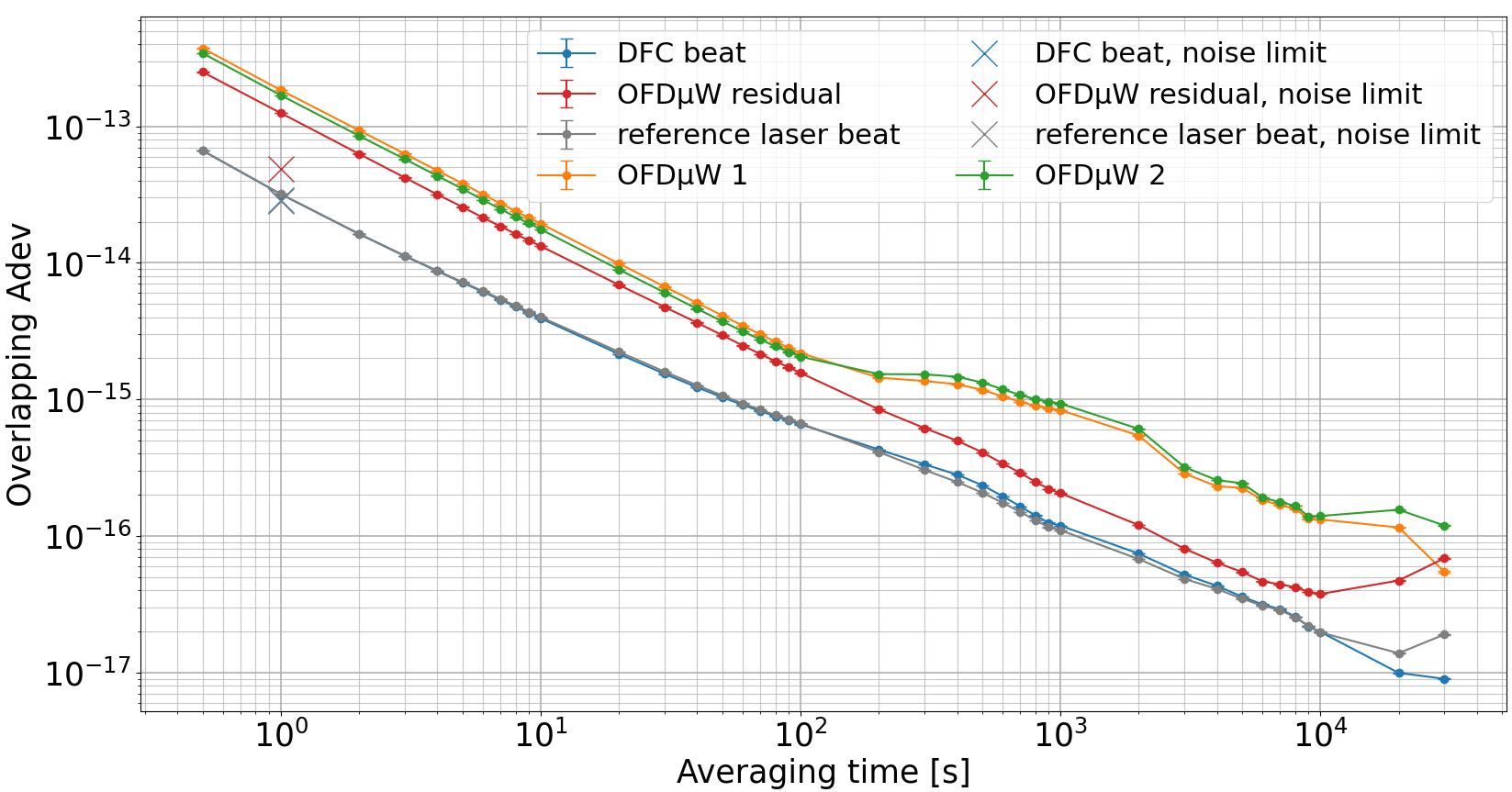}
\label{fig_ADev3}
\caption{The fractional frequency stability of the two systems.
The OADev of the individual microwaves mixed with the 9.5\,GHz LO  (orange and green), respectively. The OADev evaluated for the difference between the two-time traces, suppressing the contribution of the common 9.5\,GHz LO. The Beats between combs (blue) and reference lasers (violet), respectively. The single points show the phase noise computed from the phase noise data of the respective signals.}
\end{figure}
\indent The simultaneously counted channels are set to frequency pi-counting with a gate time of 0.5\,s. The difference between the counted microwave signals represents the stability transfer from the common 800\,MHz RF reference output to the optically generated microwaves. 
Since the 9.5\,GHz LO is used in both paths, its contribution to long-term drifts is mostly negligible for the differential between both 100\,MHz signals. 
For comparison, the optical beats between the frequency combs and the CW fiber lasers, respectively, are measured. The approximately 20’000 times higher carrier frequencies of the optical signals allow to resolve the signals with the frequency counter.\\
\indent For the optical beats between combs and reference lasers, the starting point of the overlapping Allan deviation (OADev) at small times corresponds well with the calculation from measured frequency noise according to \cite{Barnes1971}:

\begin{equation}
 \sigma_y = \int_{0}^{f_h} S_y(f) \frac{sin^4(\pi \tau f)}{(\pi \tau f)^2} \,df\ .
\end{equation}
\vspace{1mm} 

\indent The measurement of the 100\,MHz OFDµW IF is limited by the counter resolution. A possible solution to this would be to multiply the signal by a certain factor before counting. This increases the absolute frequency fluctuations of the signal and allows for slightly higher resolution by the counter for timescales up to 100\,s \cite{Lipphardt2023}. For an averaging time of 10000\,s, the fractional frequency stability of the OFDµW differential and the optical beat measurements reach a OADev level of 4e-17 and 2e-17, respectively.

\section{Conclusion}
In this work, we presented a practical implementation of optical frequency division microwave (OFDµW) generation, utilizing a CW ULN laser reference and a frequency comb stabilized to absolute frequency standards. Specifically, we demonstrated low-noise optical microwave generation based on difference-frequency generation frequency combs, where the passive elimination of the comb $f_{\mathrm{ceo}}$ enables the use of all actuators for $f_{\mathrm{rep}}$-stabilization without the risk of crosstalk. Long-term stability and absolute accuracy of the comb and OFDµW can be provided by stabilization to a 10\,MHz reference, disciplined by GPS. This setup significantly enhances the applicability of optical microwave sources in scenarios requiring a well-defined and stable oscillator in the GHz range. The resulting microwave signals were characterized in terms of pulse timing, phase noise, and fractional frequency stability. This work demonstrates the potential of DFG-combs for low-noise, high-stability microwave generation for future applications in fields such as precision timing, radar, telecommunications, optical quantum technologies, and scientific instrumentation, where both low phase noise and long-term frequency stability are essential.

\section*{References}
\bibliography{Bibliography}
\end{document}